# Translation correlations in anisotropically scattering media

Benjamin Judkewitz[1,2,*], Roarke Horstmeyer[1,*], Ivo M. Vellekoop[3],

Ioannis N. Papadopoulos[2] and Changhuei Yang[1]

1: California Institute of Technology, Departments of Electrical and Bioengineering

1200 E California Blvd, Pasadena, CA 91125

2: Exzellenzcluster NeuroCure, Charité Berlin, Humboldt University

Charitéstr. 1, 10117 Berlin, Germany

3: MIRA Institute for Biomedical Technology and Technical Medicine, University of Twente,

P.O. Box 217, 7500 AE Enschede, Netherlands

*: equal contribution

Correspondence to: benjamin.judkewitz@charite.de

# ABSTRACT

Controlling light propagation across scattering media by wavefront shaping holds great promise for a wide range of communications and imaging applications. However, finding the right wavefront to shape is a challenge when the mapping between input and output scattered wavefronts (i.e. the transmission matrix) is not known. Correlations in transmission matrices, especially the so-called memory-effect, have been exploited to address this limitation. However, the traditional memory-effect applies to thin scattering layers at a distance from the target, which precludes its use within thick scattering media, such as fog and biological tissue. Here, we theoretically predict and experimentally verify new transmission matrix correlations within thick anisotropically scattering media, with important implications for biomedical imaging and adaptive optics.

# INTRODUCTION

Focusing light through strongly scattering media is an important goal in optical imaging and communication. Long considered impossible, recent advances in the field of wavefront-shaping [1,2] changed this view by demonstrating that diffuse light can be focused through inhomogeneous media – as long as the correct input wavefront is used. With direct optical access to the target plane, the correct wavefront can be obtained by iterative optimization [2], phase-conjugation [3], or by measuring the transmission matrix [4,5]. In many imaging scenarios, however, there is no direct access to the target plane. In those cases, nonlinear [6], fluorescent [7], acousto-optic [8-10] and photo-acoustic [11-13] guide-stars can be used as reference beacons. However, these techniques only provide wavefront information for one target location at a time. While transmission matrices can be sampled quickly with a photo-acoustic approach [14], this method requires absorbing samples. As a result, many samples' transmission matrices can only be sampled sparsely. Correlations within a transmission matrix can compensate for sparse sampling and could enable high-speed imaging. One of the most widely known transmission matrix correlations is the so-called "memory effect" [15,16], which describes the following phenomenon: when an input wavefront reaching a diffusing sample is tilted within a certain angular range, the output wavefront is equally tilted, resulting in the translation of the far-field speckle pattern at a distance behind the sample (see Fig. 1).

The translation distance within which this effect holds (i.e., the field-of-view [FOV]), is inversely proportional to diffuser thickness $L$ and directly proportional to the distance $s$ of the diffuser from the screen. It can be approximated by the equation, $FOV \approx s\lambda/\pi L$ [17-20].

The memory effect has found numerous applications for point-scanning [21,22], direct image transfer [18] as well as for computational image recovery [19,20]. Yet, in all of these applications, the target plane was at a distance from a thin diffuser with free space in between ($s > 0$) – which has limited use for imaging inside thick scattering media. Because such samples are neither thin nor at a distance from the target area of interest, the correlations predicted by the "traditional" memory effect should be minimal [23]. Here, we set out to ask whether there are other correlations that apply to such samples. We show that significant transmission correlations can exist in thick scattering media at zero distance, as long as scattering is directional.

## TRADITIONAL MEMORY EFFECT

While the memory effect has been extensively derived from first principles [16], these derivations relied upon assuming perfectly diffuse scattering – which does not apply to many biological samples. Here we will approach the problem without relying upon diffusion, but instead using transmission matrices. Specifically, we are interested in the matrix $T_x = T(x_a, x_b)$, which defines the relationship between the spatial input and output optical modes of a scattering slab. For simplicity of graphical representation, we assume propagation of 1D wavefronts in a 2D geometry, but all conclusions will be generalizable to 2D wavefronts in a 3D geometry. Due to its discrete nature, the transmission matrix is especially amenable to experimental observation. We will first use our framework to analyze the traditional memory effect, and then calculate the speckle correlations in thick anisotropic media.

When assuming highly randomizing transmission, but incomplete measurement of input and output channels, the transmission matrix is often modeled as a random matrix with complex Gaussian elements. However, in the case of a thin scattering slab, the transmission matrix will have an additional macroscopic structure: a point-source on the input plane of the slab would spread to a diffuse spot at the output plane (whose diameter would be on the order of $L$ for slabs with thickness $L$ larger than one transport mean free path – as predicted by the diffusion approximation). As a result, even though individual transmission matrix elements may not be known, a finite envelope will produce a bell-shaped profile along each transmission matrix column, turning $T_x$ into a band matrix (see Figure 2a).

To recognize tilt correlations, we are interested in how the band structure of $T_x$ manifests itself in the spatial frequency domain (k-space) representation. Every spatial-domain transmission matrix $T_x$ can the transformed into the corresponding frequency-domain transmission matrix $T_k = T(k_a, k_b)$ by the following operation: $T_k = FT_xF^{-1}$, where F is a discrete Fourier transform (DFT) matrix. Due to the Fourier Inversion Theorem, we can express the inverse Fourier transform operator as the Fourier transform operator flipped: $FT_XF^{-1} \propto FT_X(F)^{\leftrightarrow} = (FT_XF)^{\leftrightarrow}$, where $\leftrightarrow$ denotes the flip operator. This is analogous to performing the 2D Fourier transform of $T_x$ and flipping it horizontally:

$$T_k \propto (\mathcal{F}_{2D}T_x)^{\leftrightarrow} \qquad (1)$$

Expressing $T_k$ in terms of the 2D Fourier transform of $T_x$ provides a straightforward explanation of how the macroscopic structure of $T_x$ influences correlations in $T_k$. Large values within $T_x$ will tend to concentrate near its diagonal, and entries will be zero elsewhere. Since $T_x$ is narrow across the diagonal, its flipped 2D Fourier transform (Eq. 1) will contain features that are elongated along the diagonal. The resulting diagonal correlations in $T_k$ correspond to the traditional memory effect, in which a tilt of the input wavefront (shift in k-space) causes a tilt of the output wavefront, resulting in a shifted speckle pattern at a distant screen. For this reason, and to distinguish the traditional memory effect from further correlations described below, we also refer to the traditional memory effect as a "tilt/tilt correlation".

The band structure of $T_x$ can be described analytically with a spatial intensity propagator matrix [24]: $P_x(x_a, x_b) \equiv \langle |T_x(x_a, x_b)|^2 \rangle$. This ensemble average removes any statistical fluctuations in $T_x$. The average spread of intensity across the output surface, from illumination with a point source input, will manifest itself in $P_x$ as a nonnegative envelope along its near-diagonals. The cross-correlation theorem may then help re-express the spatial intensity propagator matrix in the frequency domain:

$$(\mathcal{F}_{2D}^{x_b \to \Delta k_b, x_a \to \Delta k_a} P_x)^{\leftrightarrow}$$

$$= \sum_{k_a, k_b} \langle T_k(k_a, k_b) T_k^*(k_a - \Delta k_a, k_b - \Delta k_b) \rangle \quad (2)$$

$$\propto C_k(\Delta k_a, \Delta k_b)$$

Here, $C_k(\Delta k_a, \Delta k_b)$ is the tilt/tilt correlation function, while $\Delta k_a = k_a - k_a{'}$ and $\Delta k_b = k_b - k_b{'}$ denote shifts in input and output wavevectors, respectively. Typically, it is assumed that, $P_x$ only depends on the difference between the input and output spatial coordinates (shift-invariance), Eq. (2) simplifies to,

$$C_k(\Delta k_a, \Delta k_b) \propto \delta_{\Delta k_a, \Delta k_b} \mathcal{F}^{\Delta x \to \Delta k_b} P_x(\Delta x), \quad (3)$$

where $\Delta x = x_b - x_a$ and $\delta$ is a discrete delta function. Eq. (3) is the well known memory effect [24]. When considering intensity transmission, Eq. (3) corresponds to the lowest-order $C^{(1)}$ term[16]. Unlike such prior work Eq. (2) also describes the general case of tilt/tilt correlations for a geometry that is not shift-invariant.

These considerations reconfirm our expectation that the traditional memory effect may be minimal in thick biological media: First, the average spread of intensity from the input to output surface will increase with sample thickness, $L$ [25]. A wider $P_x(\Delta x)$ will subsequently reduce the range of tilt/tilt correlations between input and output plane. Second, the plane of interest is not at a distance from the sample, which means that the tilt at the output plane would not translate into a useful spatial shift at the target plane.

## CORRELATIONS IN ANISOTROPIC MEDIA

We therefore asked whether there might be other types of transmission matrix correlations in thick samples, such as biological media. We started by recognizing that in many samples scattering is anisotropic and occurs primarily in the forward direction. Scattering is particularly anisotropic in biological media, where the anisotropy parameter $g$ (the average cosine of the scattering angle) typically ranges from 0.9 to 0.98 [26,27]. This means that after a limited number of scattering events, the directionality of an input beam will be preserved to some extent as it reaches the output plane. In other words, one input plane wave (one mode in k-space) will result in a limited angular span of output waves.

As a result of such preserved directionality, anisotropically scattering media will have a macroscopic structure in the $T_k$ matrix (rather than $T_x$). Large amplitudes in $T_k$ will primarily concentrate near its main diagonal (see Fig. 2b). By analogy to our prior reasoning for the traditional memory effect, if $T_k$ is a band matrix, the entries within each diagonal of its associated $T_x$ will be correlated with one another. Now, a spatial *shift* of the optical field at the input plane will cause a spatial *shift* of the field at the output plane (this is in contrast to the traditional memory effect, in which a tilt at the input plane causes a tilt at the output plane).

The range of correlations in $T_x$ will depend upon the width of the diagonally bordered envelope in $T_k$. Equivalent to Eq. 2, we may define a k-space intensity propagator using an ensemble average: $P_k(k_a, k_b) \equiv \langle |T_k(k_a, k_b)|^2 \rangle$. Here, the average intensity spread of each input plane wave into a finite output wavevector "cone" now specifies the envelope shape along $P_k$'s rows. Following Eq. 3, we may show that the band structure of $P_k$ creates correlations in space:

$$(\mathcal{F}_{2D}^{k_b \to \Delta x_b, k_a \to \Delta x_a} P_k)^\leftrightarrow$$
$$= \sum_{x_a, x_b} \langle T_x(x_a, x_b) T_x^*(x_a - \Delta x_a, x_b - \Delta x_b) \rangle \quad (4)$$
$$\propto C_x(\Delta x_a, \Delta x_b)$$

with $C_x(\Delta k_a, \Delta k_b)$ the shift/shift correlation function. If the k-space propagator $P_k$ only depends upon difference coordinates, then Eq. (4) reduces to,

$$C_x(\Delta x_a, \Delta x_b) \propto \delta_{\Delta x_a, \Delta x_b} \mathcal{F}^{\Delta k \to \Delta x_b} P_k(\Delta k), \quad (5)$$

where $\Delta k = k_b - k_a$. This result predicts the existence of shift/shift correlations that are the exact Fourier conjugate of the traditional (tilt/tilt) memory effect.

In Eq. 5, $P_k$ includes the effects of sample anisotropy. Its width scales inversely proportional with $g$ but will increase with sample thickness $L$ (see Supplementary Material C). If $P_k$ only depends upon difference coordinates, the correlation function can be predicted by a simple experiment, namely by illuminating the sample with a plane wave $E_p(x_a) = 1$ and measuring the output wavefront, $E_p(x_b) = T_x E_p(x_a)$. The spectrum of $E_p(x_b)$ indicates the average spread of wavevectors exiting the output surface:

$$P_k(\Delta k) \equiv \left| \mathcal{F}^{\Delta x_b \to \Delta k} E_p(x_b) \right|^2. \qquad (6)$$

The Fourier transform of Eq. (6) subsequently yields the correlation between the electric field outputs from two spatially shifted inputs. In effect, the shape of the shift/shift correlation function equals the autocorrelation of $E_p(x_b)$. A more detailed derivation of Eq. (1)-(6) may be found in the Supplementary Material.

## EXPERIMENTAL VALIDATION

In order to validate our predictions experimentally, we created four scattering samples with well-defined scattering properties (3 μm diameter silica beads dispersed in agarose gel, g = 0.978 as calculated by Mie theory and $1/\mu_s$ = 175 μm at 632 nm, slab thickness $L$ in μm: 140, 280, 560 and 1120, or 1,2,4 and 8 spacers of 140 μm thickness). We then performed four different experiments with this scattering sample set, as detailed below.

First, we illuminated each sample with a diffuse input wave and recorded the output wave, $E_0$. We translated each sample laterally (Δx ranging from – 10 μm to 10 μm in 2 μm steps) and measured the correlation C(Δx) between the resulting output wavefronts, $E_{\Delta x}$, and $E_0$: C(Δx) = corr($E_0$, $E_{\Delta x}$) (Figure 3 a and c). Second, to compare these results with our predictions in Eq. (6), we illuminated each of the samples with a plane wave and calculated the autocorrelation of the output speckle patterns (Figure 3 b and d). Third, our theory predicts that the correlation function for one sample thickness $L_0$ can be used to estimate the correlation function for any other thickness (see Eq. C3 in the Supplementary Material); e. g. the correlation function for a slab of thickness $2L_0$ is simply the correlation function for a slab of thickness $L_0$, squared. We therefore used the speckle autocorrelation measured for the thinnest slab (blue curve in Figure 3d) and calculated the remaining correlation functions using Eq. C4 in the Supplementary Material (Figure 3e). Fourth, we used Mie theory to obtain a single scattering phase function for our bead samples, and calculated the theoretical correlation function using Eq. C1 in the Supplementary Material (Figure 3f). We computed this last set of curves using only the refractive indices of the media, the bead diameter and slab optical thickness (i.e. without using experimental data), as detailed in Supplementary Material C. Figure 3 f combines all four strategies for determining the correlation function into one plot. It illustrates that the experimentally measured correlation function is in agreement with all three of our derived predictions.

The shift/shift correlations apply to any input field, including fields that are shaped to converge to a sharp focus. To demonstrate the use of these correlations for scanning a point focus across a biological sample, we first used optical phase conjugation [3] to focus light (780 nm diode laser) through 500 μm to 1 mm thick slices of chicken muscle tissue, employing off-axis holography for wavefront measurement and a spatial-light-modulator (SLM) for wavefront shaping (Fig. 4a) [10,28]. We projected a point-source at one surface of the tissue slice (surface A) and detected the scattered wavefront propagating from this point through the tissue (exiting at surface B) to the SLM-plane. In the next step, we displayed the phase-conjugate of the detected wavefront, which travelled back through the tissue and formed a focus on surface A.

To validate the predicted shift/shift correlations, we then shifted the phase-conjugated wavefront laterally at surface B, testing whether the focus at surface A would be preserved and whether it moved. As expected, we noticed that motion of the shaped wavefront resulted in concurrent movement of the focus (Figure 4b-c), while the focus intensity decreased with distance from the original position, following a bell-shaped curve (Figure 4d-e). For the 500 μm slice the full width at half of the maximum (FWHM) was 5 μm, while the full-width at tenth of the maximum (FWTM) was 10 μm. In the case of the 1000 μm slice, the FWHM was 3 μm and the FWTM was 6 μm.

Published scattering parameters for chicken tissue vary, and due to experimental limitations, the single-scattering phase function has not been determined. However, Eq. (6) provides a practical way to predict the shift/shift correlation function from the experimentally accessible speckle autocorrelation function. We therefore illuminated the samples with a plane wavefront and asked whether the shape of the spatial autocorrelation of the resulting speckle pattern followed the profile of the shift/shift correlations ($C_x$), as derived in Eq. (6). Indeed, Figure 4d-e shows that both profiles are in good experimental agreement.

# DISCUSSION

The traditional (tilt/tilt) memory effect has recently enabled the development of several modalities to image through scattering 'walls' [18-22]. Intriguing as these methods are, they suffer from two limitations: the sample should be thin, and the object should be placed at a distance behind the sample.

Here, we demonstrated a complementary type of memory effect that suffers from neither limitation: the correlations are present even inside thick scattering media, as long as scattering is anisotropic and the mapping between input and output wavefronts preserves any level of directionality. This is the case up to a depth of about one transport mean free path.

We showed that the shift/shift memory effect is the Fourier complement of the traditional (tilt/tilt) memory effect, and that the extent of correlations can be directly determined from the spatial speckle autocorrelation function during plane-wave illumination.

Our results pave the way for extending memory-effect-based imaging methods [18-22] to also work inside biological tissue. Based on our measurements, we expect such methods to achieve diffraction-limited resolution at a depth of 1 mm inside muscle tissue, albeit at a limited field-of-view of <10 μm, initially.

We foresee several possibilities to further increase the field-of-view of our method, including tiling neighboring fields-of-view using multiple corrections. Also, our results suggest that the extent of correlations will be largest for photons that have undergone few scattering events and little angular deviation – also called snake-photons. Hence, selective measurement and correction of snake-photons (e.g. by temporal gating, coherence gating or spatial filtering) may considerably increase the extent of correlations and the imaging field-of-view.

Finally, we note that tilt/tilt and shift/shift correlations are not mutually exclusive. Furthermore, we anticipate that there may be additional correlations present in biological media. Future work measuring complete transmission matrices in the adaptive optics and the complex wavefront shaping regime will shed light on spectral, temporal and spatial correlations. They may ultimately be utilized in combination with the shift/shift correlations reported here.

We note that the described shift/shift correlations are consistent with the setup geometry of adaptive optics microscopy (Fig 5), where wavefronts are corrected in the conjugate Fourier plane of the microscope objective.

Tilting the incoming wavefront in the Fourier plane (e.g. in a laser scanning microscope) leads to a shift of the wavefront reaching the sample with a resulting shift of the focus [29]. In other words, adaptive optics microscopy implicitly already takes advantage of shift/shift correlations, albeit in the ballistic regime – as such it can be interpreted as a special case of the general shift/shift correlations derived here.

With further study of spatial, spectral and temporal transmission matrix correlations, these advances may lead to a unified understanding of adaptive optics and complex wavefront shaping and extend their use in thick biological tissues, enabling versatile imaging and photostimulation in a wide range of biologically relevant media.

## ACKNOWLEDGMENTS

We thank Richard Chen for providing very helpful feedback on this manuscript. This work was supported by the German Research Foundation, DFG (EXC 257 NeuroCure), NIH 1DP2OD007307-01 and the Wellcome Trust (B.J.).

## AUTHOR CONTRIBUTIONS

B.J. and R.H. conceived and developed the idea with essential help from I.M.V. B.J. and I.N.P. performed experiments. B.J., R.H., and I.M.V. wrote the manuscript. R.H. and I.M.V. wrote the mathematical supplement with help from B.J. B.J. and C.Y supervised the project.

# FIGURE LEGENDS

**Figure 1 | The traditional memory effect**. (a) the traditional memory effect as described for light propagating through thin diffusing slabs. Tilting the input wavefront (plane A) reaching the slab tilts the scattered wavefront at the output (plane B), which shifts the far-field intensity speckle pattern projected on a screen (plane C). (b) when the input wavefront is shaped to converge at a spot on the screen, tilting the input wavefront scans the spot laterally, which can be used for imaging by point scanning. The field-of-view of this approach is approximated by the equation $FOV \approx s\lambda/\pi L$.

**Figure 2 | Correlations within transmission matrices (simulated).** (a) the traditional (tilt/tilt) memory effect explained in terms of transmission matrix correlations. A pencil beam illuminating a thin slab, will cause a diffuse spot at the output surface, whose diameter $d_X$ is on the order of the slab thickness $L$. The profile of the spot will be apparent in the (ordered) X/X transmission matrix, and results in strong near-diagonal components and zeros elsewhere. The corresponding K/K transmission matrix is diagonally smeared (since X/X and K/K transmission matrices are related by the 2D Fourier transform). Hence, a tilt (k-shift) at the input plane results in a corresponding tilt (k-shift) at the output plane. (b) in anisotropically scattering media of finite thickness, the directionality of the input light may be preserved. As a result, a plane wavefront illuminating the sample creates a limited spread of output wavefronts, $d_K$. This suggests that the K/K transmission matrix of anisotropic samples has elements of higher magnitude near its diagonal. This results in a diagonally smeared X/X transmission matrix, indicating shift/shift correlations.

**Figure 3 | Experimental validation** (a) Experimental setup for determining shift/shift correlation, C(x), directly (b) Experimental setup for determining C(x) using the speckle autocorrelation resulting from plane-wave illumination (c) Experimentally measured C(x) using the setup in a. (d) C(x) obtained from the speckle autocorrelations measured in b. (e) Correlation functions calculated from the speckle autocorrelation for the thinnest sample (blue line in panel d) (f) Correlation functions predicted by the radiative transfer equation (RTE) using a phase function obtained by Mie theory. (g) overlay

**Figure 4 | Using shift/shift correlations for focusing** (a) Experimental setup (b) the time-reversed spot (middle) and shifted foci resulting from laterally shifting the phase conjugated wavefront at the sample. (c) line

scan (intensity profile) along the blue dotted line in b while shifting the input wavefront, d and e: focus peak intensity as a function of shifted location for 500/1000 μm (d/e) slices. Black curve: prediction based on the speckle autocorrelation measured during plane-wave illumination (Eq. 2).

**Figure 5 | Comparison between correlations and relation to adaptive optics** (a) the traditional memory effect, in which a tilt at the input leads to a tilt at the output face of the sample. (b) shift/shift correlations, in which a point can be scanned by shifting the corrected input wavefront. (c) typical setup in adaptive optics microscopy, where corrected wavefronts are tilted in the Fourier-plane. This leads to a shift at the sample surface, analogous to the correlations described here.

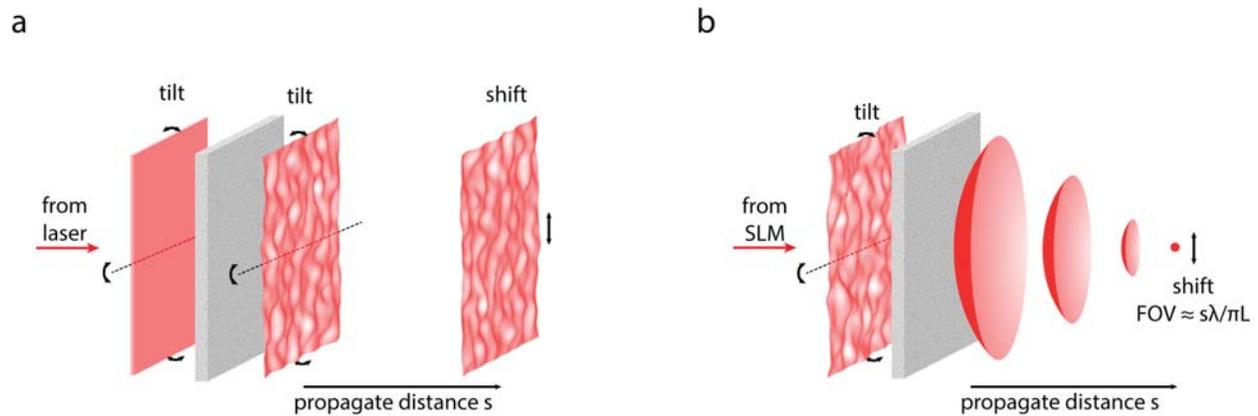

**Figure 1 | The traditional memory effect.** (a) the traditional memory effect as described for light propagating through thin diffusing slabs. Tilting the input wavefront (plane A) reaching the slab tilts the scattered wavefront at the output (plane B), which shifts the far-field intensity speckle pattern projected on a screen (plane C). (b) when the input wavefront is shaped to converge at a spot on the screen, tilting the input wavefront scans the spot laterally, which can be used for imaging by point scanning. The field-of-view of this approach is approximated by the equation FOV ≈ sλ/πL

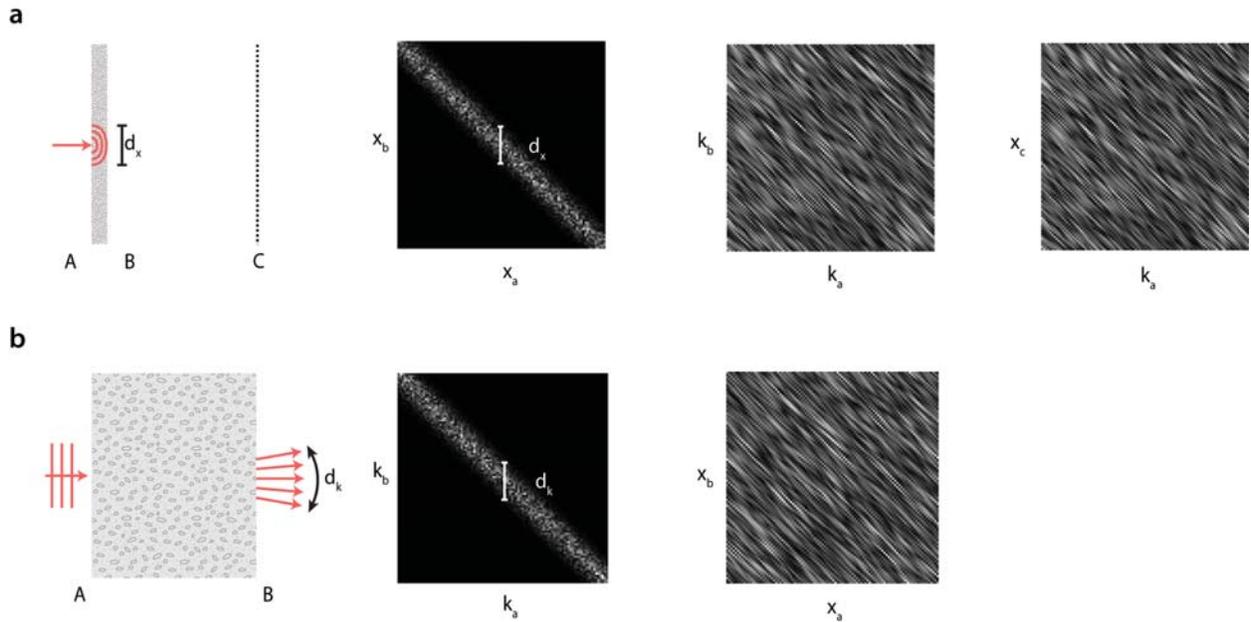

**Figure 2 | Correlations within transmission matrices (simulated).** (a) the traditional (tilt/tilt) memory effect explained in terms of transmission matrix correlations. A pencil beam illuminating a thin slab, will cause a diffuse spot at the output surface, whose diameter dX is on the order of the slab thickness L. The profile of the spot will be apparent in the (ordered) X/X transmission matrix, and results in strong near-diagonal components and zeros elsewhere. The corresponding K/K transmission matrix is diagonally smeared (since X/X and K/K transmission matrices are related by the 2D Fourier transform). Hence, a tilt (k-shift) at the input plane results in a corresponding tilt (k-shift) at the output plane. (b) in anisotropically scattering media of finite thickness, the directionality of the input light may be preserved. As a result, a plane wavefront illuminating the sample creates a limited spread of output wavefronts, dK. This suggests that the K/K transmission matrix of anisotropic samples has elements of higher magnitude near its diagonal. This results in a diagonally smeared X/X transmission matrix, indicating shift/shift correlations.

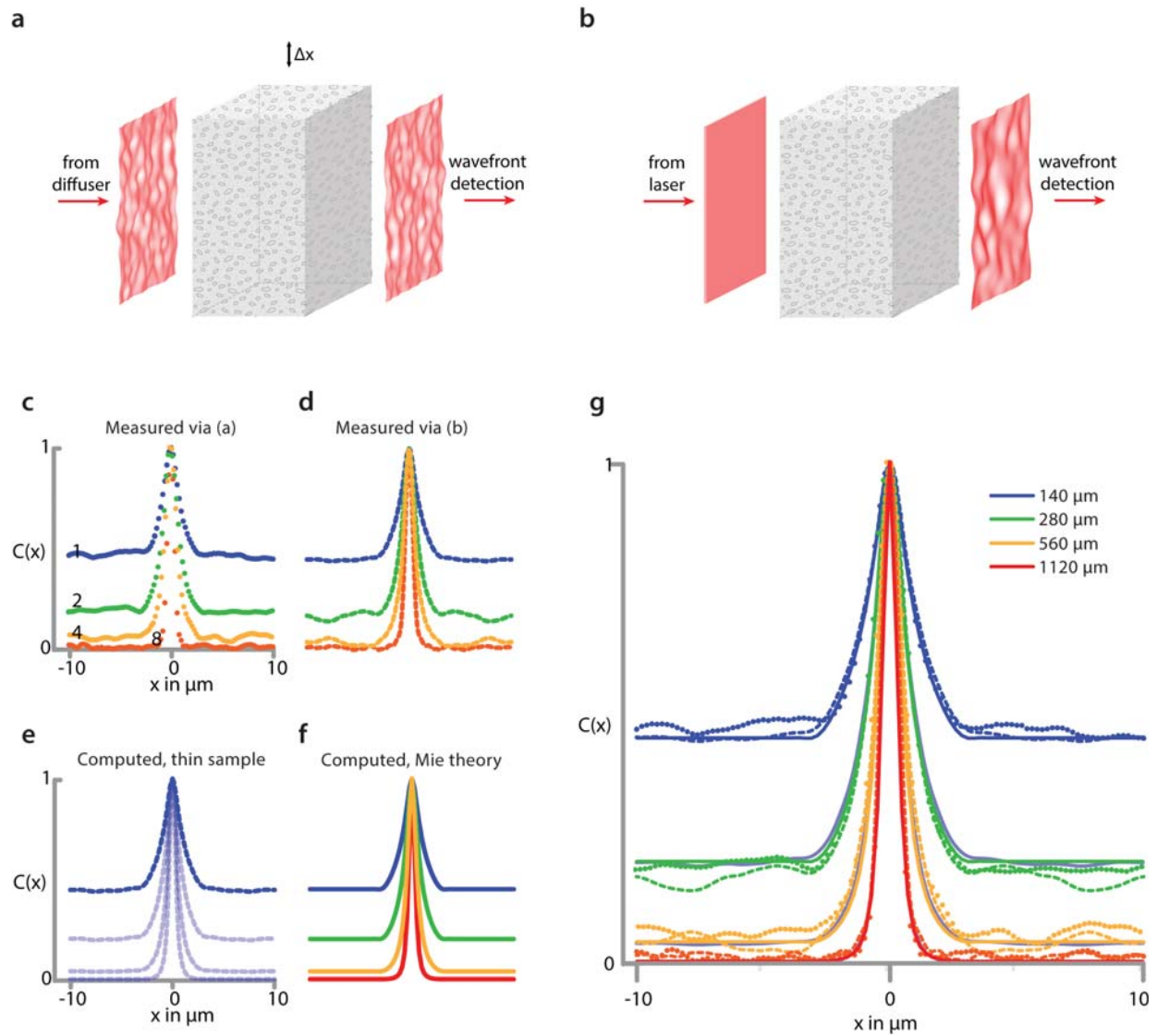

**Figure 3 | Experimental validation** (a) Experimental setup for determining shift/shift correlation, C(x), directly (b) Experimental setup for determining C(x) using the speckle autocorrelation resulting from plane-wave illumination (c) Experimentally measured C(x) using the setup in a. (d) C(x) obtained from the speckle autocorrelations measured in b. (e) Correlation functions calculated from the speckle autocorrelation for the thinnest sample (blue line in panel d) (f) Correlation functions predicted by the radiative transfer equation (RTE) using a phase function obtained by Mie theory. (g) overlay

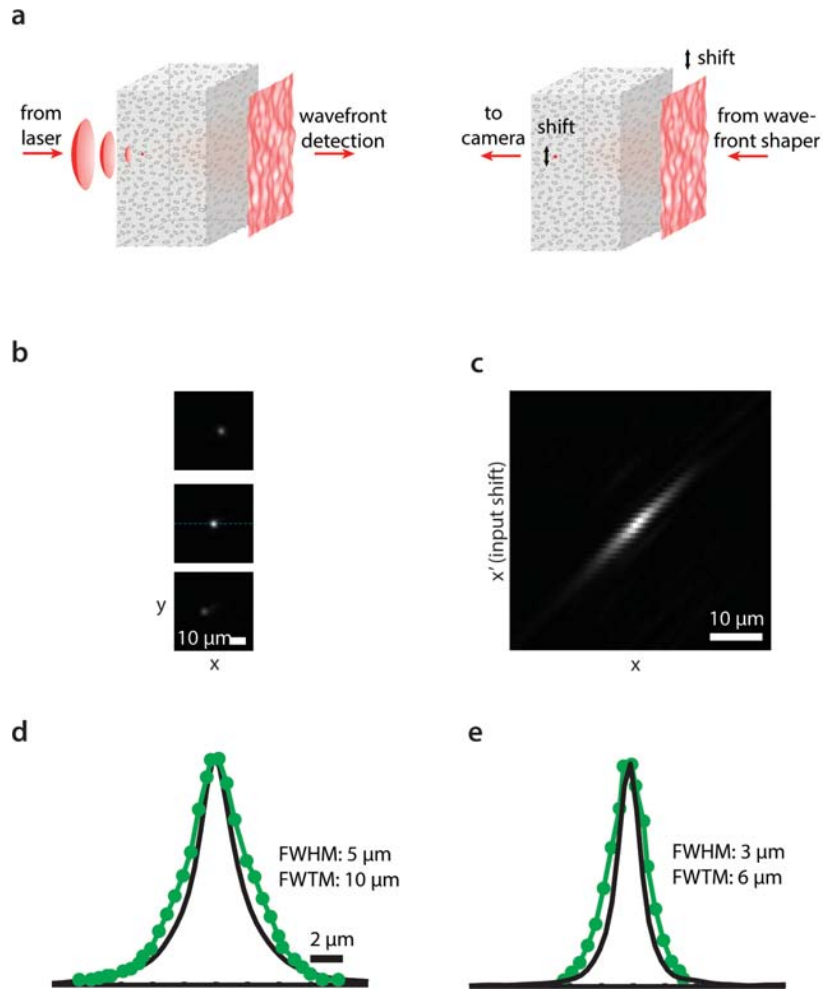

**Figure 4 | Using shift/shift correlations for focusing** (a) Experimental setup (b) the time-reversed spot (middle) and shifted foci resulting from laterally shifting the phase conjugated wavefront at the sample. (c) line scan (intensity profile) along the blue dotted line in b while shifting the input wavefront, d and e: focus peak intensity as a function of shifted location for 500/1000 μm (d/e) slices. Black curve: prediction based on the speckle autocorrelation measured during plane-wave illumination (Eq. 2).

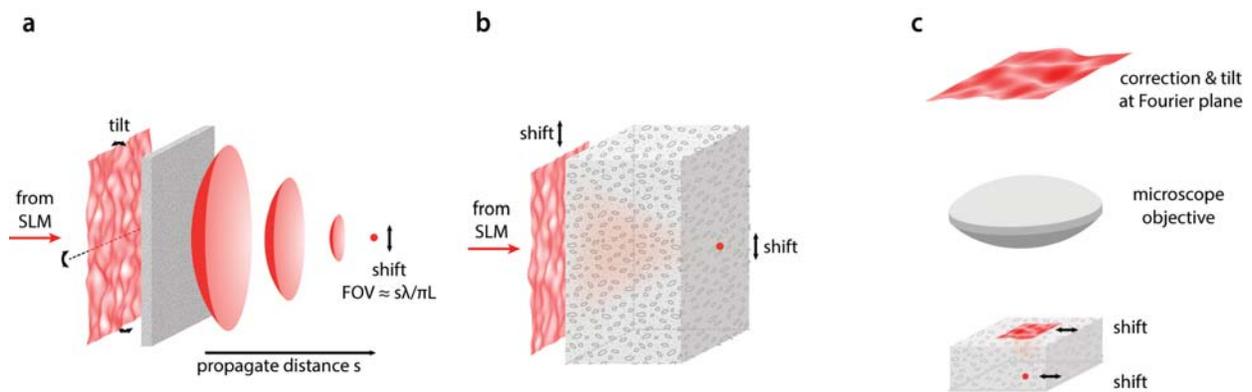

**Figure 5 | Comparison between correlations and relation to adaptive optics** (a) the traditional memory effect, in which a tilt at the input leads to a tilt at the output face of the sample. (b) shift/shift correlations, in which a point can be scanned by shifting the corrected input wavefront. (c) typical setup in adaptive optics microscopy, where corrected wavefronts are tilted in the Fourier-plane. This leads to a shift at the sample surface, analogous to the correlations described here.

# Supplementary material for: Translation correlations in anisotropically scattering media

## Supplementary material A: Rigorous derivation of the anisotropic memory effect

Here we present a rigorous derivation of the anisotropic memory effect. We prove that the anisotropic memory effect is present for all scattering objects, including ordered or absorbing ones, as long as 1) light propagation is linear, and 2) the directionality of the incident beam is maintained to any extent (i.e., the k-space intensity propagator $P_k$ is not constant across angles).

For generality, we consider the case of continuous fields $E(x_a)$ and $E(x_b)$ along the front and back surfaces of a scattering slab. We discretize this analysis in Supplementary Material B. Wave propagation through any linear medium can be described by a complex transmission function $T_x(x_a, x_b)$ such that,

$$E(x_b) = \int T_x(x_a, x_b) \, E(x_a) \mathrm{d}^2 x_a. \tag{A1}$$

Here, $E(x_a)$ is the incident field on side A, $E(x_b)$ is the transmitted field on side B, $x_a$ and $x_b$ are two-dimensional spatial coordinates in arbitrary planes in front of and behind the sample, respectively.

We now proceed to define the k-space intensity propagator, $P_k(k_b, k_a)$. The k-space intensity propagator gives the average transmitted intensity $I(k_b)$ when the medium is illuminated by a plane wave (with unit power) on side A. By convention, we define the intensity as a function of propagation direction $k_b$ as, $I(k_b) \equiv |E(k_b)|^2$.

First, we construct a truncated plane wave with unit power:

$$E_A(x_a) = \frac{H_A(x_a)}{\sqrt{A}} e^{ik_a \cdot x_a}, \quad H_A(x_a) \equiv \begin{cases} 1 & \text{for } x_a \text{ inside a square area } A \\ 0 & \text{otherwise} \end{cases} \tag{A2}$$

Here, we have $\int_A |E_A(x_a)|^2 \mathrm{d}^2 x_a = 1$. Note that while Eq. (A2)'s wave is truncated to a square area $A$, we will take the limit of $A \to \infty$ below. Following Eq. (A1), the incident field $E_A(x_a)$ will generate the following transmitted field:

$$E_A(x_b) = \frac{1}{\sqrt{A}} \int T_x(x_a, x_b) H_A(x_a) e^{ik_a \cdot x_a} \, \mathrm{d}^2 x_a. \tag{A3}$$

The resulting k-space intensity propagator is thus defined as the average spectrum intensity at side B, when an incident plane wave with wavevector $k_a$ illuminates side A:

$$P_k(k_b, k_a) \equiv \lim_{A \to \infty} \left\langle \left| \int e^{-ik_b \cdot x_b} E_A(x_b) \, \mathrm{d}^2 x_b \right|^2 \right\rangle \tag{A4}$$

$$= \left\langle \frac{1}{A} \iint e^{-ik_b \cdot x_b} T_x(x_a, x_b) H_A(x_a) e^{ik_a \cdot x_a} \mathrm{d}^2 x_a \mathrm{d}^2 x_b \right.$$

$$\left. \times \iint e^{ik_b \cdot x_{b'}} T_x^*(x_{a'}, x_{b'}) H_A(x_{a'}) e^{-ik_a \cdot x_{a'}} \, \mathrm{d}^2 x_{a'} \mathrm{d}^2 x_{b'} \right\rangle$$

Here, the averaging is performed over many random scatterer configurations. It is direct to show that this definition of $P_k$ is equivalent to the form used in the main text. Next, we apply the coordinate transform $x_{a'} \to x_a - \Delta x_a$, $x_{b'} \to x_b - \Delta x_b$ and reorder Eq. (A4) to find,

$$P_k(k_b, k_a) = \lim_{A \to \infty} \frac{1}{A} \iiiint \langle T_x(x_a, x_b) T_x^*(x_a - \Delta x_a, x_b - \Delta x_b) \rangle \quad (A5)$$

$$\times H_A(x_a) H_A(x_a - \Delta x_a) e^{ik_a \cdot \Delta x_a} e^{-ik_b \cdot \Delta x_b} d^2 x_a \, d^2 x_b \, d^2 \Delta x_a \, d^2 \Delta x_b$$

We now define our shift-shift correlation function $C_x$ in the limit $A \to \infty$:

$$C_x(\Delta x_a, \Delta x_b) \equiv \lim_{A \to \infty} \frac{1}{A} \iint H_A(x_a) H_A(x_a - \Delta x_a) \quad (A6)$$

$$\times \langle T_x(x_a, x_b) T_x^*(x_a - \Delta x_a, x_b - \Delta x_b) \rangle \, d^2 x_a \, d^2 x_b,$$

$$= \lim_{A \to \infty} \frac{1}{A} \int_A \int_A \langle T_x(x_a, x_b) T_x^*(x_a - \Delta x_a, x_b - \Delta x_b) \rangle \, d^2 x_a \, d^2 x_b.$$

Note that Eq. (A6)'s correlation function can always be defined, even when the medium is not shift-invariant. However, defining it only makes sense when the medium is statistically invariant to translations over the area of interest. We may now insert Eq. (A6) into Eq. (A5) to find,

$$P_k(k_b, k_a) = \iint C_x(\Delta x_a, \Delta x_b) e^{ik_a \cdot \Delta x_a} e^{-ik_b \cdot \Delta x_b} d^2 \Delta x_a \, d^2 \Delta x_b. \quad (A7)$$

Inverting this Fourier transform gives,

$$C_x(\Delta x_a, \Delta x_b) = \iint P_k(k_b, k_a) e^{-ik_a \cdot \Delta x_a} e^{ik_b \cdot \Delta x_b} d^2 k_a \, d^2 k_b, \quad (A8)$$

which is the continuous equivalent to Eq. (4) in the main manuscript. For convenience, we can change the coordinates of $P_k$ to $k \equiv k_a$ and $\Delta k \equiv k_b - k_a$ and write,

$$C_x(\Delta x_a, \Delta x_b) = \iint P_k'(k; \Delta k) e^{i\Delta k \cdot \Delta x_b} e^{-ik \cdot (\Delta x_a - \Delta x_b)} d^2 k \, d^2 \Delta k, \quad (A9)$$

which is still a general expression. Note that in most circumstances $C_x(\Delta x_a, \Delta x_b) \neq 0$, even for $\Delta x_a \neq \Delta x_b$.

**Interpretation**

In its most general form, Eq. (A9)'s shift-shift correlation function depends on both $\Delta x_a$ and $\Delta x_b$. If we want to calculate the magnitude of the anisotropic memory effect, we evaluate the correlation function at $\Delta x_a = \Delta x_b = \Delta x$ to find that it is simply the Fourier transform of an angle-averaged angular intensity propagator:

$$C_x(\Delta x) = \int \overline{P_k}(\Delta k) e^{i\Delta k \cdot \Delta x} d^2 \Delta k. \quad (A10)$$

Here, $\overline{P_k}$ is an angle-averaged k-space intensity propagator that we define as, $\overline{P_k}(\Delta k) \equiv \int P_k'(k; \Delta k) d^2 k$. The width of $\overline{P_k}(\Delta k)$ depends on the scattering object's mean free path, anisotropy coefficient, thickness, reflections at the object surface, and possibly other parameters. If a disordered scattering object's width $L$ is much thicker than the transport mean free path for light $l$, all directionality is lost and $\overline{P_k}(\Delta k)$ will be a constant function across $\Delta k$. In this case, the width of $\overline{P_k}(\Delta k)$ will still be limited to $2k_0$, with $k_0 \equiv 2\pi/\lambda$. This finite support of $\overline{P_k}$ gives rise to trivial correlations on the scale of half a wavelength.

If a disordered scattering object's width $L$ is thinner than one transport mean free path $l$, some level of directionality is preserved and $\overline{P_k}(\Delta k)$ will be narrower than for the case when $L \gg l$. The narrower the angular intensity propagator's support, the larger the distance over which the sample can be translated while preserving correlations in the speckle at the scatterer's back surface.

**Special case: $P_k$ is separable**

If Eq. (A9)'s intensity propagator $P_k(k; \Delta k)$ is separable into $P_k^0(k)$ and $P_k^\Delta(\Delta k)$, we may rewrite it in the form,

$$C_x(\Delta x_a, \Delta x_b) = \int P_k^\Delta(\Delta k)\, e^{i\Delta k \cdot \Delta x_b} \mathrm{d}^2\Delta k \int P_k^0(k) e^{-ik \cdot (\Delta x_b - \Delta x_a)}\, \mathrm{d}^2 k, \quad (A11)$$

which is the product of the Fourier transform of $P_k^\Delta$ with the inverse Fourier transform of $P_k^0$. In the special case where $P_k$ only depends on the difference between the incident and transmitted wave angle $\Delta k$, we find that $P_k^0$ is constant. Under this condition, Eq. (A11) simplifies to,

$$C_x(\Delta x_a, \Delta x_b) = \delta(\Delta x_b - \Delta x_a) \int P_k(\Delta k) e^{i\Delta k \Delta x_b}\, \mathrm{d}^2\Delta k. \quad (A12)$$

In Eq. (A12)'s final form, it is immediately clear that the anisotropic memory effect is the real-space analogy to the traditional memory effect, matching Eq. (5) in the main text.

**Conclusion**

The anisotropic memory effect exists *for any linear medium*, as long as any level of directionality is maintained. The effect of shifting the medium is always robustly defined as the Fourier transform of the k-space intensity propagator, $\overline{P_k}(\Delta k)$. In the $C_1$ approximation, the intensity-intensity correlation function is simply obtained by taking the absolute square of the field-field correlation $C_x$ in Eq. (A11) or Eq. (A12). Finally, the interested reader may replace all variables and functions above with their respective Fourier conjugates. An analogous, generalized derivation of the traditional memory effect will be the result.

**Supplementary Material B: Discretization into transmission matrices**

It is direct to transfer our mathematical findings for the continuous case in Supplementary Material A to the discrete equations introduced in the main text.

First, we may apply Shannon's sampling theorem to discretize the input and output fields, $E(x_a)$ and $E(x_b)$, into vectors. Assuming a 1D geometry, we set the resolution of each vector element at $\delta x_a = \delta x_b = \lambda/2NA$, where the numerical aperture $NA$ is the maximum acceptance angle of light on the input/output surface (which we here assume are equal, for simplicity). Given the area of the input and output surfaces each equal $A$, then we find the number of spatial modes on each surface as $n = 2NA \cdot A/\lambda$. We may write $E_A(x_a)$ and $E_A(x_b)$ [from Eq. (A2)-(A3)] each as 1×$n$ vectors, which are now connected by a discrete $n$×$n$ transmission matrix, $T_x(x_a, x_b)$.

We may now re-write Eq. (A4) in discrete form as,

$$P_k(k_b) \equiv \langle |\mathcal{F}_{1D}^{x_b \to k_b} E_A(x_b)|^2 \rangle \tag{B1}$$

where $\mathcal{F}$ is a discrete Fourier transform operator and $P_k(k_b)$ is now one column of our discrete k-space intensity propagator matrix. Eq. (A1) connects $E_A(x_b)$ to the spatial transmission matrix $T_x$. Thus, we may generate multiple averaged fields, each under plane wave illumination from a unique wavevector $k_a$, and place their Fourier transform in one row of a matrix:

$$P_k(k_a, k_b) \equiv \langle |T_k(k_a, k_b)|^2 \rangle. \tag{B2}$$

Eq. (B2) is the k-space intensity propagator matrix defined in the main text.

**Supplementary Material C: Theoretical prediction of the k-space intensity propagator**

Here, we detail one way to connect the anisotropic memory effect's k-space intensity propagator, $P_k(k_a, k_b)$, to scattering material parameters like $g$ and $L$.

Prior derivations of the traditional memory effect (e.g., [Li1994], [Ber1989]) use the radiative transport equation (RTE) in the diffusion approximation to connect the intensity propagator, $P_x(x_a, x_b)$, to scattering material parameters. The diffusion approximation typically requires a material thickness of several transport mean-free paths. The anisotropic memory effect will be most pronounced, and of greatest use, within one transport mean-free path. Thus, we opt here to follow a different derivation route for the anisotropic memory effect. Our alternative solution more accurately models the transport of light through thin, anisotropic slabs.

Instead of using the diffusion approximation to simplify the RTE, [Ish1978, Kok1997] present an alternative approximation based upon the small-angle approximation (SAA). This limit assumes all light is highly forward scattered (g > 0.8, approximately). Many biological materials of interest, including the tissue tested in our experiment, satisfy this condition.

If we assume the optical setup of interest is cylindrically symmetric, [Kok1997] uses the SAA of the RTE to predict the intensity spectrum across the back surface of a scatterer, when under plane wave illumination, $I_k(k_b)$:

$$I_k(k_b) = \frac{1}{\lambda} \cdot \mathcal{F}^{x \to k_b} \exp\left[-\tau\left(1 - \omega \hat{h}(x)\right)\right]. \tag{C1}$$

Here, $\omega$ is the scattering material albedo, $\tau = \alpha L$ is its optical thickness, $\alpha$ is the scattering cross section, and $\hat{h}(x)$ is the Fourier transform of the single scattering phase function: $\hat{h}(x) = \mathcal{F}^{\theta \to x} h(\theta)$. For example, $h(\theta)$ may take the form of the well known Henyey-Greenstein phase function:

$$h(\theta) = \frac{1 - g^2}{(1 + g^2 - 2g \cos\theta)^{3/2}}, \tag{C2}$$

where $g$ is the anisotropy (asymmetry) parameter. Alternatively, Mie theory can lead to and exact solution for $h(\theta)$. For example, to create the curve in Figure 3f, we insert the closed-form Mie scattering solution of $h(\theta)$ for a homogeneous sphere (from [Boh1983]) into Eq. (C1).

If we assume the k-space intensity propagator is shift-invariant in k-space, then we may express its one-dimensional form [see Eq. (4) and Eq. (A12)] as, $P_k(\Delta k) = I_k(k_b)$. Shift invariance is not required to connect the SAA with the anisotropic memory effect, but greatly simplifies all relations. Eq. (C1)-Eq. (C2) now directly connect the anisotropic memory effect, through the Fourier transform of the k-space intensity propagator $P_k(\Delta k)$, to material parameters like optical thickness ($\tau$) and anisotropy ($g$).

A multiplicative relationship between sample optical thickness $\tau$ and the shift/shift correlation function $C_x$ directly follows from Eq. (C1). For a slab of thickness $2\tau$, we first assume shift-invariance to write, $P_k(\Delta k, 2\tau) \propto \mathcal{F}^{x \to \Delta k} \exp\left[-2\tau\left(1 - \omega \hat{h}(x)\right)\right]$. This can be equivalently expressed as the Fourier transform of a product: $P_k(\Delta k, 2\tau) \propto \mathcal{F}^{x \to \Delta k}[\hat{P}_k(x,\tau) \cdot \hat{P}_k(x,\tau)]$. Here, $\hat{P}_k(x,\tau)$ is the Fourier transform of the intensity propagator for a slab of thickness $\tau$: $\hat{P}_k(x,\tau) = \mathcal{F}^{\Delta k \to x} P_k(\Delta k, \tau)$. Since our shift/shift correlation function is also the Fourier transform of $P_k$, we thus conclude that,

$$P_k(\Delta k, 2\tau) \propto \mathcal{F}^{\Delta x \to \Delta k}[C_x(\Delta x, \tau) \cdot C_x(\Delta x, \tau)]. \tag{C3}$$

Equivalently, we may remove the Fourier transform to find,

$$C_x(\Delta x, 2\tau) = C_x(\Delta x, \tau) \cdot C_x(\Delta x, \tau). \tag{C4}$$

We apply the multiplicative relationship in Eq. (C4) to compute the curve shown in Figure 3e of the main text.

References

[Li1994] J. H. Li and A. Z. Genack, *Phys. Rev. E* **49**, 4530 (1994).

[Ber1989] R. Berkovitz, M. Kaveh, & S. Feng, "Memory effect of waves in disordered systems: A real-space approach", *Physical Review B* 40(1), p.737-740 (1989).

[Ish1978] A. Ishimaru, *Propagation and Scattering in Random Media: Volume 1* (New York: Academic Press, 1978), Ch. 13.

[Kok1997] A. Kokhanovsky, "Small angle approximations of the radiative transfer theory," *J Phys D: Appl Phys* 30, 2837-2840 (1997).

[Boh1983] C. F. Bohren and D. R. Huffman, *Absorption and scattering of light by small particles* (New York: Wiley, 1983), Ch. 4.
6